\pdfoutput=1
\documentclass[useAMS,usenatbib,usegraphicx]{mn2e}
\usepackage{times}
\usepackage{color}
\usepackage{aas_macros}

\newcommand{\kms}{km s\ensuremath{^{-1}}}

\title[Fourier Band-Power E/B-mode Estimators for Cosmic Shear]{Fourier Band-Power E/B-mode Estimators for Cosmic Shear}
\author[M. R. Becker \& E. Rozo]{Matthew R. Becker$^{1,2,}$\thanks{E-mail: beckermr@stanford.edu} and Eduardo Rozo$^{2,3}$\\
$^{1}${KIPAC, Physics Department, Stanford University, Stanford, CA 94305} \\
$^{2}${KIPAC, SLAC National Accelerator Laboratory, Menlo Park, CA 94025}\\
$^{3}$Department of Physics, University of Arizona, Tucson, AZ 85721}

\begin{document}

\date{}

\pagerange{\pageref{firstpage}--\pageref{lastpage}} \pubyear{2014}

\maketitle

\label{firstpage}

\begin{abstract}
We introduce new Fourier band-power estimators for cosmic shear data analysis and E/B-mode separation. 
We consider both the case where one performs E/B-mode separation and the case where one does not. The resulting 
estimators have several nice properties which make them ideal for cosmic shear data analysis. First, they can be 
written as linear combinations of the binned cosmic shear correlation functions. Second, they account for the survey window 
function in real-space. Third, they are unbiased by shape noise since they do not use correlation function data at zero separation. 
Fourth, the band-power window functions in Fourier space are compact and largely non-oscillatory. Fifth, they can be used to construct 
band-power estimators with very efficient data compression properties. In particular, we find that all of the information on the parameters $\Omega_{m}$, 
$\sigma_{8}$ and $n_{s}$ in the shear correlation functions in the range of $\sim10-400$ arcminutes for single tomographic bin can be compressed into only three
 band-power estimates. Finally, we can achieve these rates of data compression while excluding small-scale information 
where the modeling of the shear correlation functions and power spectra is very difficult. Given these desirable properties, 
these estimators will be very useful for cosmic shear data analysis.
\end{abstract}


\section{Introduction}\label{sec:intro}
\setcounter{footnote}{0}

Cosmic shear, or the weak gravitational lensing of background galaxies by large-scale structure, is one of the most 
promising cosmological probes because it can in principle provide direct constraints on the amplitude and shape of the 
projected matter power spectrum. Through measuring the growth of the matter power spectrum, cosmic shear will 
provide strong constraints on the nature Dark Energy \citep[e.g.,][]{weinberg2013}, interesting parameters like neutrino masses 
\citep[e.g.,][]{kitching2008} and new ideas like modifications to General Relativity \citep[e.g.,][]{beynon2010}. Motivated by the intrinsic 
statistical power of cosmic shear measurements, a large number of surveys are either ongoing or have been planned in 
part to take cosmic shear measurements, including the 
DES\footnote{The Dark Energy Survey - http://www.darkenergysurvey.org}, 
HSC\footnote{Hyper Suprime-Cam - http://www.naoj.org/Projects/HSC}, 
KIDS\footnote{The Kilo Degree Survey - http://kids.strw.leidenuniv.nl}, 
Pan-STARRS\footnote{The Panoramic Survey Telescope \& Rapid Response System - http://pan-starrs.ifa.hawaii.edu}, 
Euclid\footnote{http://sci.esa.int/euclid}, 
LSST\footnote{Large Synoptic Survey Telescope - http://www.lsst.org} and 
WFIRST\footnote{Wide-Field Infrared Survey Telescope - http://wfirst.gsfc.nasa.gov}.

It is expected that these cosmic shear experiments will be difficult, being subject to many potential systematic effects 
in both the measurements and the modeling \citep[see, e.g.,][for a review]{weinberg2013}. Cosmic shear measurements are made 
by correlating the lensed shapes of galaxies with one another. As galaxies are approximately, 
but not exactly \citep[see, e.g.,][for a review]{troxel2014}, randomly oriented in the absence of lensing, we can attribute large-scale correlations among the 
galaxy shapes to gravitational lensing. However, we observe galaxies through the atmosphere and telescope which change their shapes 
through the point spread function (PSF). These instrumental effects can potentially be much bigger than the signals we are looking for and can mimic true 
cosmic shear signals. Thus they must be removed carefully. 

Luckily, cosmic shear has several built-in null tests than can be used to search for and verify the absence of contamination 
in the signals. Checking for B-mode contamination in the cosmic shear signals is one of the most important of these null tests \citep{kaiser1992}.
Weak gravitational lensing at the linear level only produces parity-free E-mode shear patterns. Small amounts of shear 
patterns with net handedness, known as B-mode patterns, can be produced by higher-order corrections, but their amplitude is generally much 
too small be observed by current surveys \citep[e.g.,][]{krause2010}. Thus we can use the absence or presence of B-mode patterns in the observed shear field to look 
for systematic errors. PSF patterns generally have similar levels of E- and B-modes unlike true cosmic shear signals. Note that ensuring the 
level of B-modes in a survey is consistent with zero is a necessary but not sufficient condition for the shear measurements to be error free. 
Other tests, such as correlating the measured shears with the PSF itself, may provide a more powerful check on the level of systematic 
contamination \citep{heymans2012}.

The importance of checking cosmic shear signals for B-mode contamination has motivated a large amount of work on devising statistical 
measures of the B-mode contamination \citep[e.g.,][]{schneider1998,seljak1998,hu2001b,schneider2002b,schneider2007,schneider2010,hikage2011,becker2013a}. 
The main obstacle confronting every B-mode estimator is the mixing of E/B-modes in the 
estimator and the effect of ambiguous modes. This mixing happens on large-scales when one considers instead of an infinitely large survey, 
a survey of finite size. For a finite sized survey, modes with wavelengths of order the patch size can sometimes not be uniquely classified 
as either E- or B-modes \citep[e.g.,][]{bunn2003}. These ambiguous modes can contaminate the E- and B-mode estimators. If all of the power in the survey is sourced 
by E-modes, then the ambiguous modes are actually E-modes which then leads to mixing of E-modes into B-modes. Note also that masking, 
binning and pixelization can also produce E- to B-mode mixing \citep[e.g.,][]{smith2006,lin2012,becker2013a}.

In this work, we present extensions to the estimators derived in \citet[][B13 hereafter]{becker2013a}. The estimators presented in B13 use linear combinations 
of the binned real-space shear two-point correlation function measurements over a finite angular range excluding zero (see below) to produce 
statistics which optimally separate E- and B-modes. They additionally have the interesting property that given any linear combination of just the 
$\xi_{+}$ correlation function (see the definitions below), one can derive through a simple linear algebra the corresponding linear combination for 
$\xi_{-}$ which maximizes the E- and B-mode separation. We will exploit this freedom to derive linear combinations of the binned real-space 
correlation function measurements which are local in Fourier space and optimally separate E- and B-modes. These Fourier space band-powers 
are competitive with more sophisticated pseudo-$C_{\ell}$ methods, producing as much or more separation between E- and B-modes ($\sim$4 orders of magnitude) 
while retaining the main advantage of shear correlation function measurements -- accounting for the survey mask in real space. 

Additionally, we derive below estimators which do not attempt to separate E- and B-modes, but instead assume all of the power is sourced by E-modes. 
They again use linear combinations of the binned shear correlation function data points to estimate a Fourier-space band-power. As the B-mode null test 
may not be the most powerful test of systematic contamination in cosmic shear measurements, we expect that these estimators will be useful 
for the analysis of cosmic shear data which has negligible systematic contamination. 

We note that the band-powers presented in this work use apodized windows in Fourier space and explicitly account for the binning of the cosmic shear 
correlation functions, unlike those presented in \citet{asgari2014}. These features decrease the spillage and ringing of the band-powers and make the analysis 
exact in the sense that the estimators account for the full effects of the survey window function. Due to the fact that the band-power window functions in Fourier 
space can be computed exactly with the formalism presented in this work, they thus have no biases which impact a cosmological analysis. One simply uses 
the correct window functions to do the theoretical predictions and compute covariances. 

Finally, we show that the estimators presented in this work can be used to construct band-powers with very efficient data compression properties, 
while remaining reasonably well-localized in both real and Fourier space. We find that all of the information on the parameters $\Omega_{m}$, $\sigma_{8}$ and 
$n_{s}$ in the shear correlation functions from an angular range of $\sim10-400$ arcminutes can be compressed into three band-powers. This conclusion is similar 
to those in \citet{asgari2014}. Performing cosmological parameter analysis in the data compressed space of band-powers can ease the requirements for generating large 
numbers of simulations for estimating covariances. While we have considered the data compression for only a single tomographic bin, further compression with our 
methods when doing full tomography should be possible as well. 

This work is organized as follows. First we given an overview of the formalism of B13 and derive the new estimators in Section~\ref{sec:est}. Then we 
give two examples of the estimators and considerations for constructing them in Section~\ref{sec:ex}. The data compression properties of the 
estimators are discussed in Section~\ref{sec:dc}. Finally, we present discussion and conclusions in Section~\ref{sec:conc}.

\section{Optimal Estimators for Input Band Windows}\label{sec:est}
In this section, we use the methods of B13 to construct estimators for the lensing 
power spectra which are linear combinations of the shear two-point correlation functions. 
We consider two cases. First, we consider the case where no E/B-mode separation is performed. In this case, one 
is assuming explicitly that the weak lensing data is free of systematic errors (or at least systematic errors which would 
cause B-modes). Second, we consider the case where E/B-mode separation is performed, but one still wants an estimate 
of the power spectrum constructed from the two-point shear correlation functions. Note that estimators derived below are very 
similar to those presented in \citet{schneider2002a} and the formalism below is very similar to that presented in \citet{asgari2014}. 
However, in this work we treat the binning of the cosmic shear correlation functions explicitly and further present apodized 
band-powers that exhibit less ringing than those presented in \citet{asgari2014}. 

Before, considering the two cases outlined above, we review key parts of the formalism of B13. Suppose one 
constructs estimators $X_{\pm}$ from the shear two-point functions as follows,
\begin{equation}
X_{\pm} = \frac{1}{2}\sum_{i}F_{+i}\hat\xi_{+i} \pm F_{-i}\hat\xi_{-,i}\ ,
\end{equation}
with 
\begin{equation}
\hat\xi_{\pm i} = \int_{L_{i}}^{H_{i}}d\theta\, W_{i}(\theta)\xi_{\pm}(\theta)\ .
\end{equation}
Here the window functions, $W_{i}(\theta)$, are assumed to be normalized to unity over the angular range considered, 
$\theta\in[L_{i},H_{i}]$. See B13 for a more extensive discussion of the window functions. For purely geometric effects, note that 
$W_{i}(\theta)=2\theta/(H_{i}^{2}-L_{i}^{2})$. Finally, recall that the shear correlation functions can 
be written as \citep[cf.][]{schneider2007}
\begin{eqnarray}
\xi_{+}(\theta) &=& \int_{0}^{\infty}\frac{d\ell\,\ell}{2\pi}J_{0}(\ell\theta)\left[P_{E}(\ell) + P_{B}(\ell)\right]\\
\xi_{-}(\theta) &=& \int_{0}^{\infty}\frac{d\ell\,\ell}{2\pi}J_{4}(\ell\theta)\left[P_{E}(\ell) - P_{B}(\ell)\right]
\end{eqnarray}
where the $J_{n}(\ell\theta)$ are cylindrical Bessel functions. $P_{E}(\ell)$ and $P_{B}(\ell)$ are the E- and 
B-mode power spectra respectively.

Given the definitions above, we can write the expectation value of 
this estimator in terms of the E- and B-mode power spectra as
\begin{equation}
\left\langle X_{\pm} \right\rangle = \int_{0}^{\infty}d\ln\ell\,\left[\frac{\ell^{2}P_{E}(\ell)}{2\pi}W_{\pm}(\ell)+\frac{\ell^{2}P_{B}(\ell)}{2\pi}W_{\mp}(\ell)\right]
\end{equation}
with
\begin{eqnarray}
W_{\pm}(\ell) &=& \frac{1}{2}\sum_{i}\left(F_{+i}\int_{L_{i}}^{H_{i}}d\theta\, W_{i}(\theta) J_{0}(\ell\theta)\right.\nonumber\\
&&\ \ \ \ \ \ \ \ \ \ \ \ \ \ \ \left. \pm F_{-i}\int_{L_{i}}^{H_{i}}d\theta\, W_{i}(\theta) J_{4}(\ell\theta)\right)\ .\nonumber
\end{eqnarray}

B13 showed that the following steps lead to estimators which optimally separate E- and B-modes (where optimal here means 
that the estimators minimize E/B mixing in the root-mean-square sense for the window function $W_{-}(\ell)$; see B13).
\begin{enumerate}
\item Pick a fiducial $F_{+}$ vector. 
\item Project out, as vectors, the quantities $F_{+a}$ and $F_{+b}$ from $F_{+}$. Note that these 
vectors correspond to ambiguous modes which cannot uniquely be classified as either E- or B-modes on a finite 
patch of sky. We denote the component of $F_{+}$ orthogonal to the ambiguous modes as $\tilde{F}_{+}$.
\item Compute $F_{-}$ from $\tilde{F}_{+}$ via a vector-matrix multiplication, $F_{-}=M_{+}\tilde{F}_{+}$.
\end{enumerate}
B13 additionally demonstrated that this process can be repeated starting with $F_{-}$, with ambiguous modes 
$F_{-a}$ and $F_{-b}$ along with the matrix $M_{-}$. 

\begin{figure}
\includegraphics[width=\columnwidth]{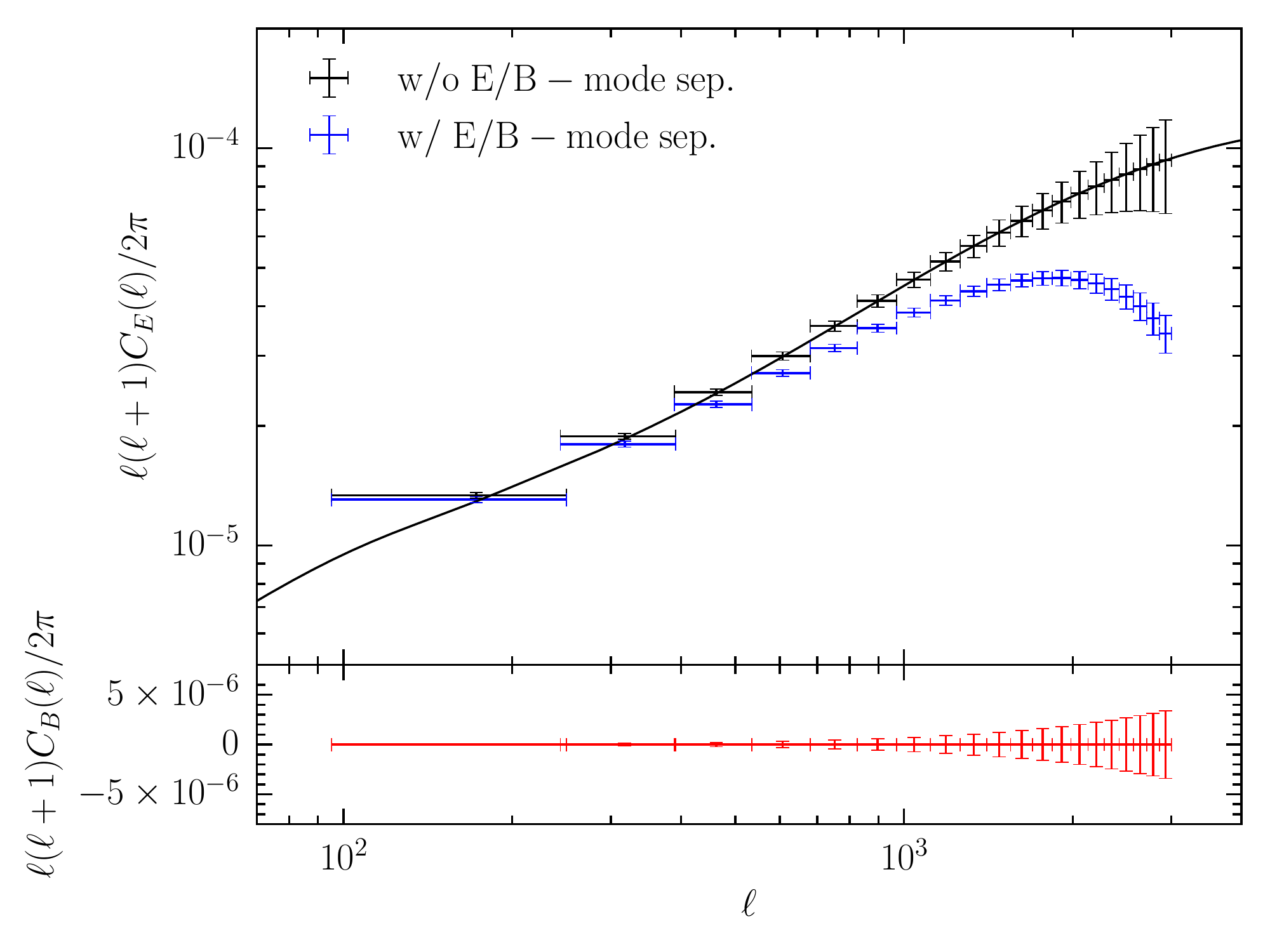}
\caption[]{Band-power estimators which do not (black points) and do (blue and red points) separate E- and B-modes. Each estimator is shown 
with x-axis error bars giving the $1\sigma$ width of the log-Normal band window function and y-axis error bars giving the $1\sigma$ 
error on the band-power amplitude for a DES-like survey assuming all sources are at redshift one and using only Gaussian covariances 
for the cosmic shear correlation functions. The top panel shows the E-mode estimators and the bottom panel shows the B-mode estimators. 
The solid line is the underlying lensing power spectrum.
\label{fig:bpow}}
\end{figure}

For completeness, we reproduce the definitions of $F_{+a}$, $F_{+b}$, $F_{-a}$, $F_{-b}$, $M_{+}$ and $M_{-}$ below. 
\begin{eqnarray}
F_{+a} &=& \left(\int_{L_{1}}^{H_{1}}d\theta\,W_{1}(\theta),\int_{L_{2}}^{H_{2}}d\theta\,W_{2}(\theta),...,\right.\nonumber\\
&&\left.\ \ \ \ \int_{L_{N}}^{H_{N}}d\theta\,W_{N}(\theta)\right)_{N}\\
F_{+b} &=& \left(\int_{L_{1}}^{H_{1}}d\theta\,W_{1}(\theta)\,\theta^{2},\int_{L_{2}}^{H_{2}}d\theta\,W_{2}(\theta)\,\theta^{2},...,\right.\nonumber\\
&&\left.\ \ \ \ \int_{L_{N}}^{H_{N}}d\theta\,W_{N}(\theta)\,\theta^{2}\right)_{N}\\
\lefteqn{F_{-a} = \left(\int_{L_{1}}^{H_{1}}d\theta\,\frac{W_{1}(\theta)}{\theta^2},\int_{L_{2}}^{H_{2}}d\theta\,\frac{W_{2}(\theta)}{\theta^2},\right.}\nonumber\\
&&\ \ \ \ \ \ \ \ \ \ \ \ \ \ \ \ \ \ \ \ \ \ \ \ \ \ \ \ \ \ \ \ \ \left. ...,\int_{L_{N}}^{H_{N}}d\theta\,\frac{W_{N}(\theta)}{\theta^2}\right)_{N}\\
\lefteqn{F_{-b} = \left(\int_{L_{1}}^{H_{1}}d\theta\,\frac{W_{1}(\theta)}{\theta^4},\int_{L_{2}}^{H_{2}}d\theta\,\frac{W_{2}(\theta)}{\theta^4},\right.}\nonumber\\
&&\ \ \ \ \ \ \ \ \ \ \ \ \ \ \ \ \ \ \ \ \ \ \ \ \ \ \ \ \ \ \ \ \ \left. ...,\int_{L_{N}}^{H_{N}}d\theta\,\frac{W_{N}(\theta)}{\theta^4}\right)_{N}\ 
\end{eqnarray}
\begin{eqnarray}
\lefteqn{(M_{+})_{ki} = \delta_{ki}  + \left(\int_{L_{k}}^{H_{k}}d\theta\frac{W_{k}^{2}(\theta)}{\theta}\right)^{-1}}&&\nonumber\\
&&\times\int_{L_{i}}^{H_{i}}\int_{L_{k}}^{H_{k}}d\theta\,d\phi\,W_{i}(\theta)\,W_{k}(\phi)\nonumber\\
&&\ \ \ \ \  \ \ \ \ \ \ \ \ \ \ \ \ \ \ \ \ \ \ \ \ \ \ \ \ \ \ \ \times\left(\frac{4}{\phi^{2}} - \frac{12\theta^{2}}{\phi^{4}}\right)H(\phi-\theta)
\end{eqnarray}
\begin{eqnarray}
\lefteqn{(M_{-})_{ki} = \delta_{ki}  + \left(\int_{L_{k}}^{H_{k}}d\theta\frac{W_{k}^{2}(\theta)}{\theta}\right)^{-1}}&&\nonumber\\
&&\times\int_{L_{i}}^{H_{i}}\int_{L_{k}}^{H_{k}}d\theta\,d\phi\,W_{i}(\theta)\,W_{k}(\phi)\nonumber\\
&&\ \ \ \ \  \ \ \ \ \ \ \ \ \ \ \ \ \ \ \ \ \ \ \ \ \ \ \times\left(\frac{4}{\theta^{2}} - \frac{12\phi^{2}}{\theta^{4}}\right)H(\theta-\phi)
\end{eqnarray}
In these expressions $H(\theta)$ is the Heaviside step function and $\delta_{ki}$ is the Kronecker delta function. See B13 for a detailed explanation of 
how these quantities are defined and derived. 

\begin{figure*}
\begin{center}
\includegraphics[width=2.2\columnwidth]{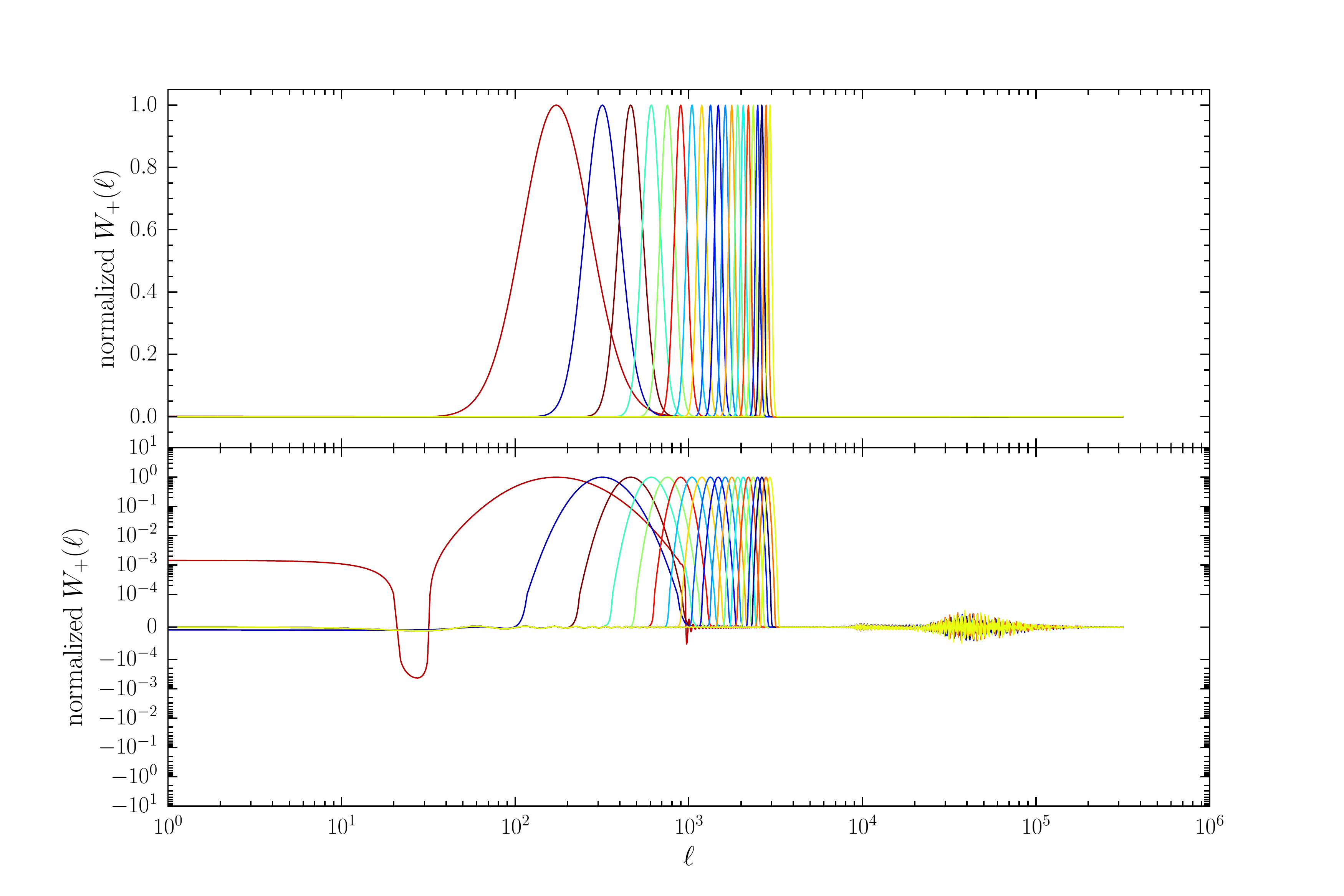}
\end{center}
\caption[]{Band-power estimator kernels $W_{+}(\ell)$ with no E/B-mode separation. Each color shows a different band-power estimators with a linear scale (top) 
and symmetric log-scale (bottom). The scaling in the bottom panel is logarithmic above/below $\pm10^{-4}$, but linear between. Each kernel has been normalized to 
unity at its peak. Note that the wings of the band-powers are both positive and negative so that under integration they largely cancel out.\label{fig:kernsnosepeb}}
\end{figure*}

\subsection{Estimators without E/B-mode Separation}
Imagine the case where one has a target band-power window, $W_{T}(\ell)$, which is used 
to produce a band-power estimate of the shear power spectrum,
\begin{equation}
D_{T} = \int_{0}^{\infty}\frac{d\ell\,\ell}{2\pi}W_{T}(\ell)P_{E}(\ell)\ .
\end{equation}
Note that for piecewise-constant power spectra, $P_{\alpha}$, over some range in $\ell\in[L_{\alpha},H_{\alpha}]$, 
this band-power is
\begin{equation}
D_{\alpha} = P_{\alpha}\int_{L_{\alpha}}^{H_{\alpha}}\frac{d\ln\ell\,\ell^{2}}{2\pi}W_{\alpha}(\ell)\ .
\end{equation}

We define estimators which optimally match the target window 
function $W_{T}(\ell)$ by minimizing the squared difference between the output window function and 
the test window function,
\begin{equation}
\int_{0}^{\infty}d\ell\,\ell \left[W_{T}(\ell)-W_{+}(\ell)\right]^{2}\ .
\end{equation}
See \citet{asgari2014} for the unbinned version of the derivation below. Minimizing the expression above 
with respect to $F_{+}$ and $F_{-}$ simultaneously (see B13 for similar algebraic manipulations), we get the following linear equations
\begin{eqnarray}
F_{+} &=& A^{0} - M_{-}F_{-}\label{eqn:fpcond}\\
F_{-} &=& A^{4} - M_{+}F_{+}\label{eqn:fmcond}\ .
\end{eqnarray}
We can solve these equations for $F_{+}$, getting
\begin{eqnarray}
(I - M_{-}M_{+})F_{+} = A^{0} - M_{-}A^{4}\ .
\end{eqnarray}
Here the $A^{0,4}$ vectors are
\begin{eqnarray}
A^{0}_{k} &=& 2\left[\int_{L_{k}}^{H_{k}}d\theta\,W_{k}^{2}(\theta)/\theta\right]^{-1}\nonumber\\
&&\ \ \ \ \ \times\int_{L_{k}}^{H_{k}}d\theta\int_{0}^{\infty}d\ell\,\ell\,W_{k}(\theta)W_{T}(\ell)J_{0}(\ell\theta)\nonumber\\
A^{4}_{k} &=& 2\left[\int_{L_{k}}^{H_{k}}d\theta\,W_{k}^{2}(\theta)/\theta\right]^{-1}\\
&&\ \ \ \ \ \times\int_{L_{k}}^{H_{k}}d\theta\int_{0}^{\infty}d\ell\,\ell\,W_{k}(\theta)W_{T}(\ell)J_{4}(\ell\theta)
\end{eqnarray}
Once $F_{+}$ is known, we can solve for $F_{-}$ trivially using Equation~(\ref{eqn:fmcond}) above. Note that the existence of this estimator requires that the matrix $I-M_{-}M_{+}$ be invertible or equivalently have an empty kernel or null space. As demonstrated in B13, for the geometric window functions defined above and logarithmic shear correlation function binning, this matrix is typically invertible (B13 tested only specific cases and did not provide a general proof). Thus we expect these estimators to exist for most surveys, but note that their detailed properties depend on the survey window functions, specific binning, etc. 

\subsection{Estimators which Separate E- and B-modes}
We define a heuristic method to construct estimators which simultaneously match 
the test window function $W_{T}(\ell)$ and separate E- and B-modes. To do this, we just optimize 
the metric above for $F_{+}$ and set $F_{-}$ by minimizing the E/B-mixing, using the steps listed above. Then we get 
\begin{equation}
\left[I+M_{-}M_{+}\left(I  - \tilde{F}_{+a}\tilde{F}_{+a}^{T} - \tilde{F}_{+b}\tilde{F}_{+b}^{T}\right) \right]F_{+} = A^{0}
\end{equation}
and $F_{-} = M_{+}\tilde{F}_{+} = M_{+}(I - \tilde{F}_{+a}\tilde{F}_{+a}^{T} - \tilde{F}_{+b}\tilde{F}_{+b}^{T})F_{+}$. Here we have 
defined projection operators for the ambiguous modes as $F_{+a}F_{+a}^{T}$ and $F_{+b}F_{+b}^{T}$. The vectors $\tilde{F}_{+a}$ and $\tilde{F}_{+b}$ 
are just the ambiguous modes, but rewritten so that they are mutually orthogonal and normalized to unity.

\begin{figure*}
\begin{center}
\includegraphics[width=2.2\columnwidth]{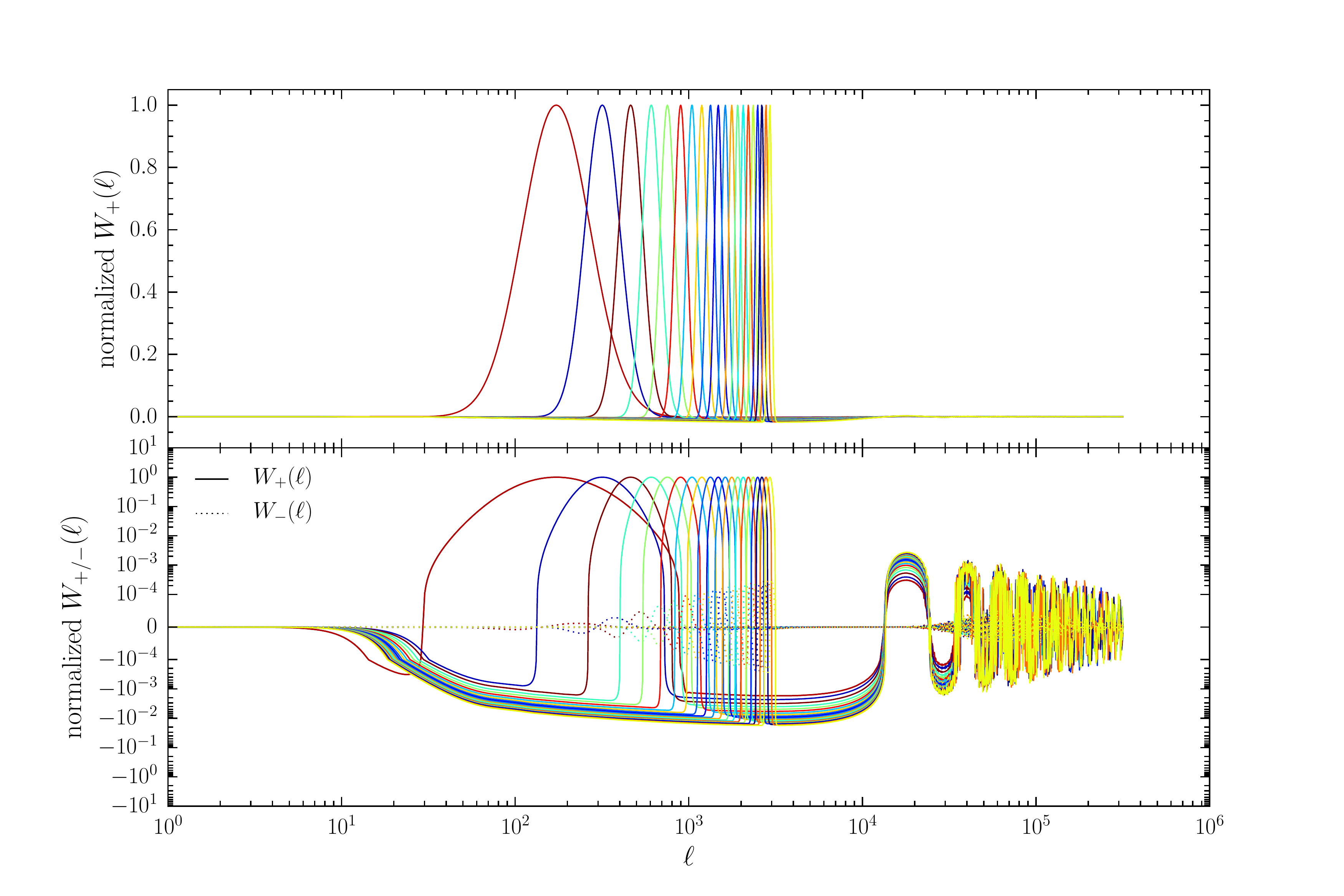}
\end{center}
\caption[]{Band-power estimator kernels $W_{+/-}(\ell)$ with E/B-mode separation. Each color shows a different band-power estimators with a linear scale (top) 
and symmetric log-scale (bottom).The scaling in the bottom panel is logarithmic above/below $\pm10^{-4}$, but linear between. Each kernel has been normalized to 
unity at its peak, with the $W_{-}(\ell)$ kernels normalized by the peak of the corresponding $W_{+}(\ell)$ kernels, so that the relative amplitudes are correct. 
In the bottom panel, the dotted lines show the $W_{-}(\ell)$ filters. Mixing between E- and B-modes is controlled by the amplitude of the $W_{-}(\ell)$ kernel, 
which is suppressed by $\sim3-4$ orders of magnitude.\label{fig:kernssepeb}}
\end{figure*}

\section{Example Sets of Estimators}\label{sec:ex}
We now give two example sets of estimators in order to illustrate their performance and to further discuss 
design considerations. Specifically, we construct estimators of both types discussed above. For concreteness, 
we consider a survey which has measured shear correlation function data in $10^{3}$ angular bins between 1 and 
400 arcminutes. We also use the geometric bin window functions given above. 

\subsection{Considerations for Choosing $W_{T}(\ell)$}
In this work, we use a log-normal target window in Fourier space
\begin{equation}
W_{T}(\ell) = \frac{1}{\sigma\sqrt{2\pi}}\exp\left[-\frac{1}{2}\left(\frac{\log\ell-\log\ell_{m}}{\sigma}\right)^{2}\right]\ .
\end{equation}
We set $\ell_{m}$ to be spaced linearly, 
\begin{displaymath}
\ell_{m,i+1/2} = \Delta\ell(i+1/2) + \ell_{min}
\end{displaymath}
where $\Delta\ell = (\ell_{max}-\ell_{min})/N$, with $\sigma=\log(\ell_{m,i+1}/\ell_{m,i})/2$. We have found empirically that 
these choices for the target band windows produce locally compact estimators which minimize ringing in Fourier space. 
We use $\left\{\ell_{min},\ell_{max},N\right\}=\left\{100,3000,20\right\}$ in this work. Other target window functions, like apodized top-hat band windows 
also work well, but we do not consider those choices in this work. 

\begin{figure*}
\begin{center}
\includegraphics[width=2.2\columnwidth]{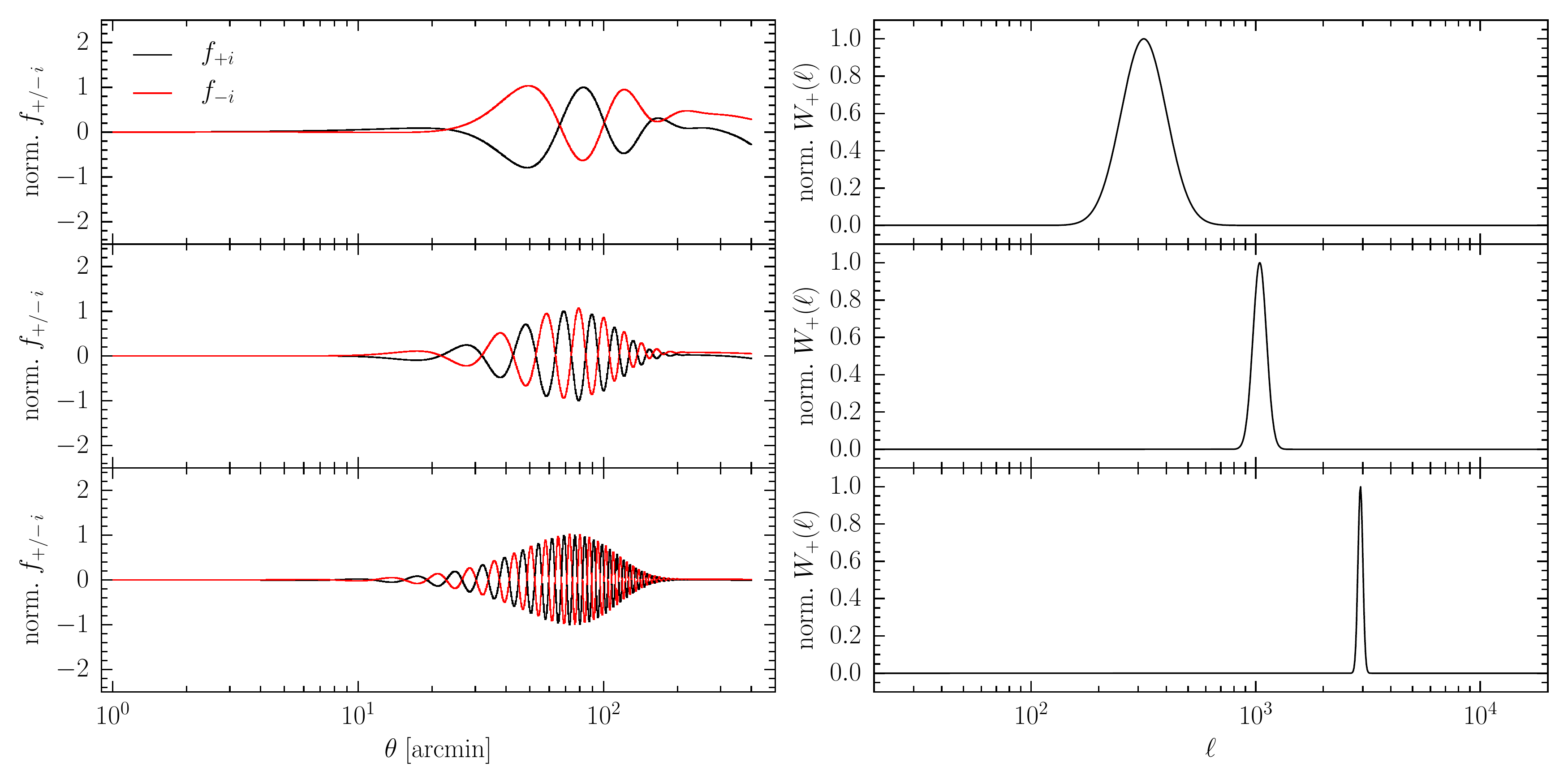}
\end{center}
\caption[]{The real-space weights $f_{+/-i}$ (left) and corresponding Fourier space window functions (right) 
for three selected band-power estimators which do not separate E- and B-modes. As expected, as the Fourier space 
window targets higher $\ell$ modes, more oscillations appear in real space. Once the real-space oscillations 
exceed the local Nyquist frequency of the real-space correlation function binning, the estimators begin to show 
significant ringing in Fourier space. \label{fig:rkernsnosepeb}}
\end{figure*}

\begin{figure*}
\begin{center}
\includegraphics[width=2.2\columnwidth]{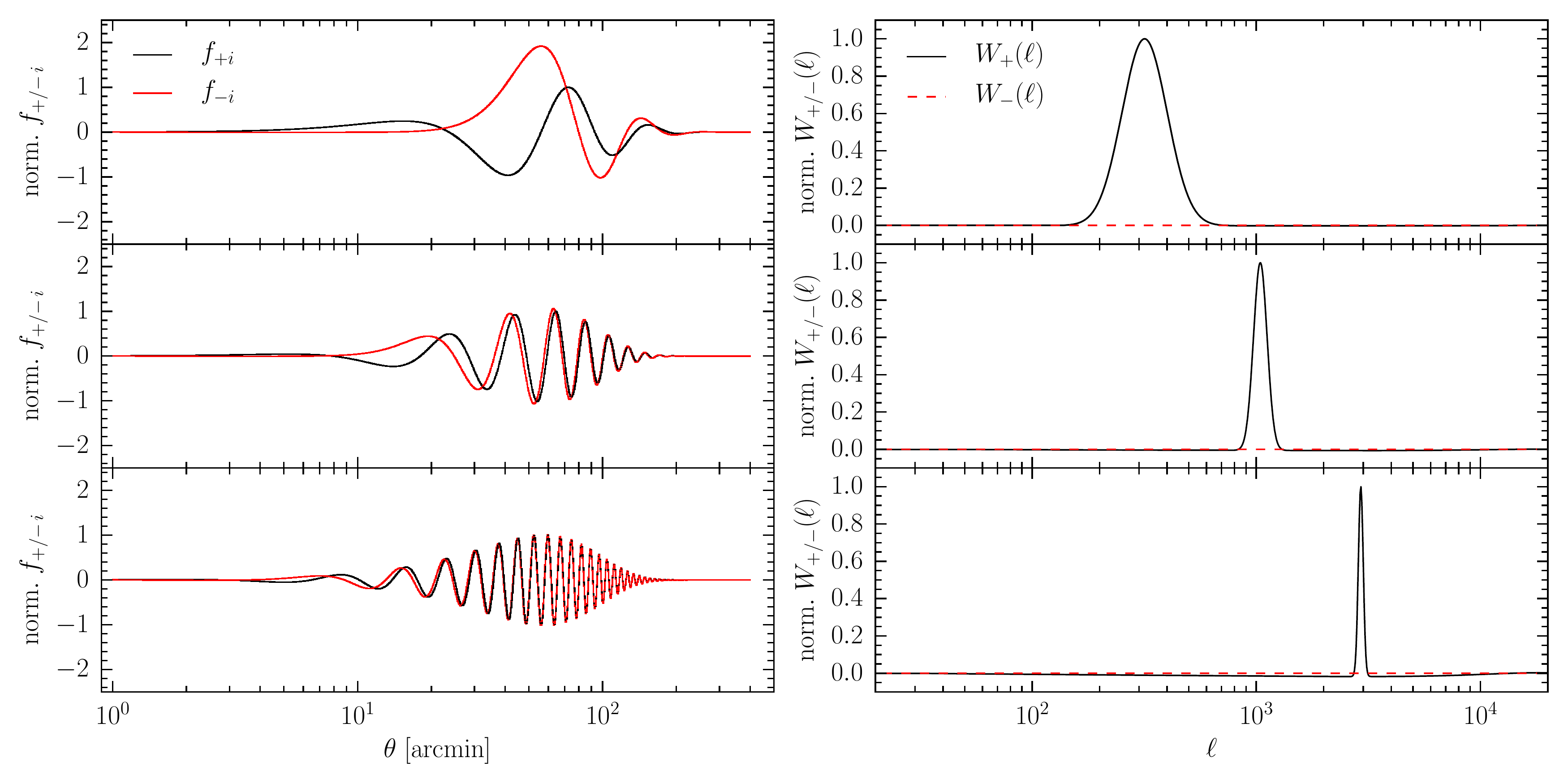}
\end{center}
\caption[]{The real-space weights $f_{+/-i}$ (left) and corresponding Fourier space window functions (right) 
for three selected band-power estimators which separate E- and B-modes. As expected, as the Fourier space 
window targets higher $\ell$ modes, more oscillations appear in real space. Once the real-space oscillations 
exceed the local Nyquist frequency of the real-space correlation function binning, the estimators begin to show 
significant ringing in Fourier space. \label{fig:rkernssepeb}}
\end{figure*}

\begin{figure*}
\begin{center}
\includegraphics[width=2\columnwidth]{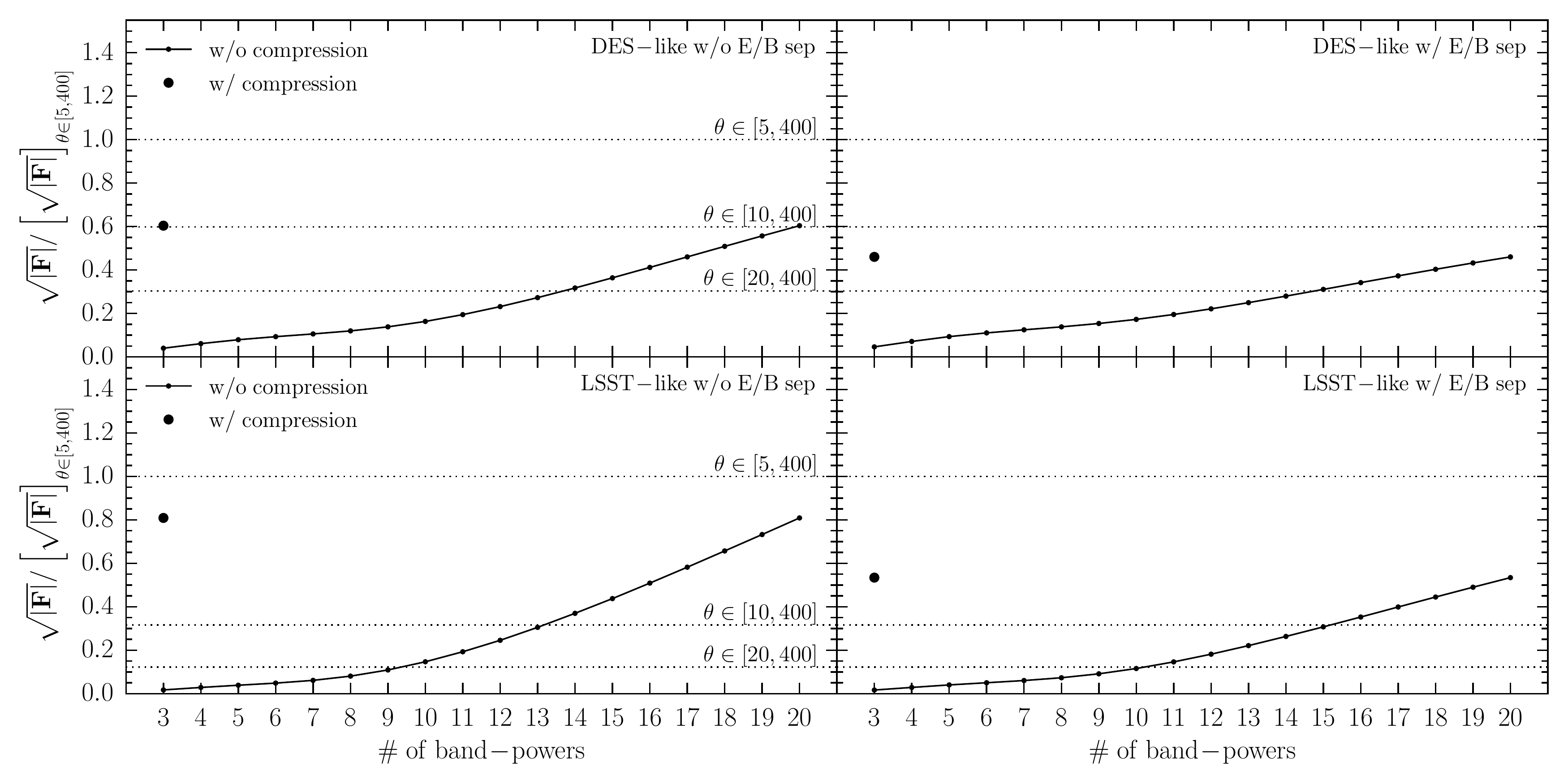}
\end{center}
\caption[]{The Fisher information content of band-power estimators for DES- (top) and LSST-like (bottom) surveys. Estimators 
without E/B-mode separation are on the left and estimators with E/B-mode separation are on the right. The solid line shows the information 
content in the case of no data compression. The information content of the compressed statistics is shown by the large solid point.  
The dotted lines from bottom top show the amount of information contained in a shear correlation function only analysis as a function of 
the smallest scale considered from 20 to 5 arcminutes. Note that the band-power estimators generaly assign very little weight to shear 
correlation function data below 10 arcminutes, but that the conversion between Fourier mode and angular scale is not exact. \label{fig:finfo}}
\end{figure*}

\subsection{The Estimators}
The band-power estimators with their expected errors, assuming Gaussian covariances for a DES-like survey (see Section~\ref{sec:dc}) 
are shown in Figure~\ref{fig:bpow}. We have assumed all sources are at redshift one for this computation. We show for each band-power the 
$1\sigma$ width of the window function and the $1\sigma$ error bar assuming the DES-like survey. The band-powers which do not 
separate E- and B-modes (black points) track the underlying power spectrum (solid line) closely. For estimators which separate E- and B-modes 
(blue points), there is a feature due to the estimator window functions at high-$\ell$. The window functions for estimators which separate E- and B-modes 
exhibit some degree of compensation (due to the removal of the ambiguous modes) and thus at high-$\ell$ deviate in the mean from the 
underlying power spectrum. Note however that this deviation can be predicted exactly so that parameter estimator done with these estimators will 
still result in unbiased parameters (assuming no other biases of course). The bottom panel of this figure shows the B-mode band-powers (red points) 
which are suppressed by $\sim4$ orders of magnitude relative to the E-mode estimators. This level of separation between the band-powers 
is competitive with standard psuedo-$C(\ell)$ methods \citep[e.g.,][]{hikage2011}.

Figures~\ref{fig:kernsnosepeb} and \ref{fig:kernssepeb} show the window functions $W_{+/-}(\ell)$ for estimators which don't or 
do separate E- and B-modes respectively. In general, the window functions are quite compact in Fourier space, but they do however 
exhibit oscillations for large $\ell$. Fortunately, the regions which exhibit these oscillations are suppressed by factor of $10^{2}$ 
to $10^{3}$ relative to the peak and under integration these oscillations will largely cancel. Additionally, for estimators which 
separate E- and B-modes, the $W_{-}(\ell)$ kernels are suppressed by $\sim3-4$ orders of magnitude, indicating that these estimators 
have a very small amount of E/B mixing.

We show the real-space filters for these estimators in Figures~\ref{fig:rkernsnosepeb} and \ref{fig:rkernssepeb} for estimators 
which don't or do separate E- and B-modes respectively. We find unsurprisingly that as the target band window function 
moves higher in $\ell$, there are more oscillations in the real-space weights $f_{+/-i}$. Additionally, we find that if 
the oscillations in the real-space weights exceed the local Nyquist frequency of the real-space correlation function binning, 
the estimators being to show significant non-locality in Fourier space. This consideration motivates the choice of the 
number of real-space bins to use to construct the estimators. We have chosen $10^{3}$ for this example, but note that for estimators 
up to $\ell\sim1000$, 500 bins in real-space works reasonably well too. Finally, note that one must have a sufficient number density of 
sources in order to use a large number of radial bins. Assuming one needs at minimum $10^{3}$ source pairs in a radial bin, then for a radial bin at 
1 arcminutes with a $\sim$0.6\% width (the width used in this work for $10^{3}$ bins in the range 1 to 400 arcminutes) for a DES-like survey, 
one requires a source density of at least $\sim0.01$ galaxies per square arcminute. All future lensing surveys clearly meet this requirement. 

Notice that in real-space, these estimators are very efficient at excluding small-scale modes in the shear correlation functions. 
Thus these estimators can be used to help render cosmic shear measurements insensitive to potential systematic effects on 
small scales, like baryonic effects in the matter power spectrum \citep[e.g.,][]{zhan2004,white2004,rudd2008,vandaalen2011,casarini2012} or 
the small-scale selection effects in shear measurements \citep[e.g.,][]{hartlap2011}. Similarly, these estimators do not use the shear correlation 
functions at zero separation, rendering them unbiased by shape noise. 

\begin{figure*}
\begin{center}
\includegraphics[width=2\columnwidth]{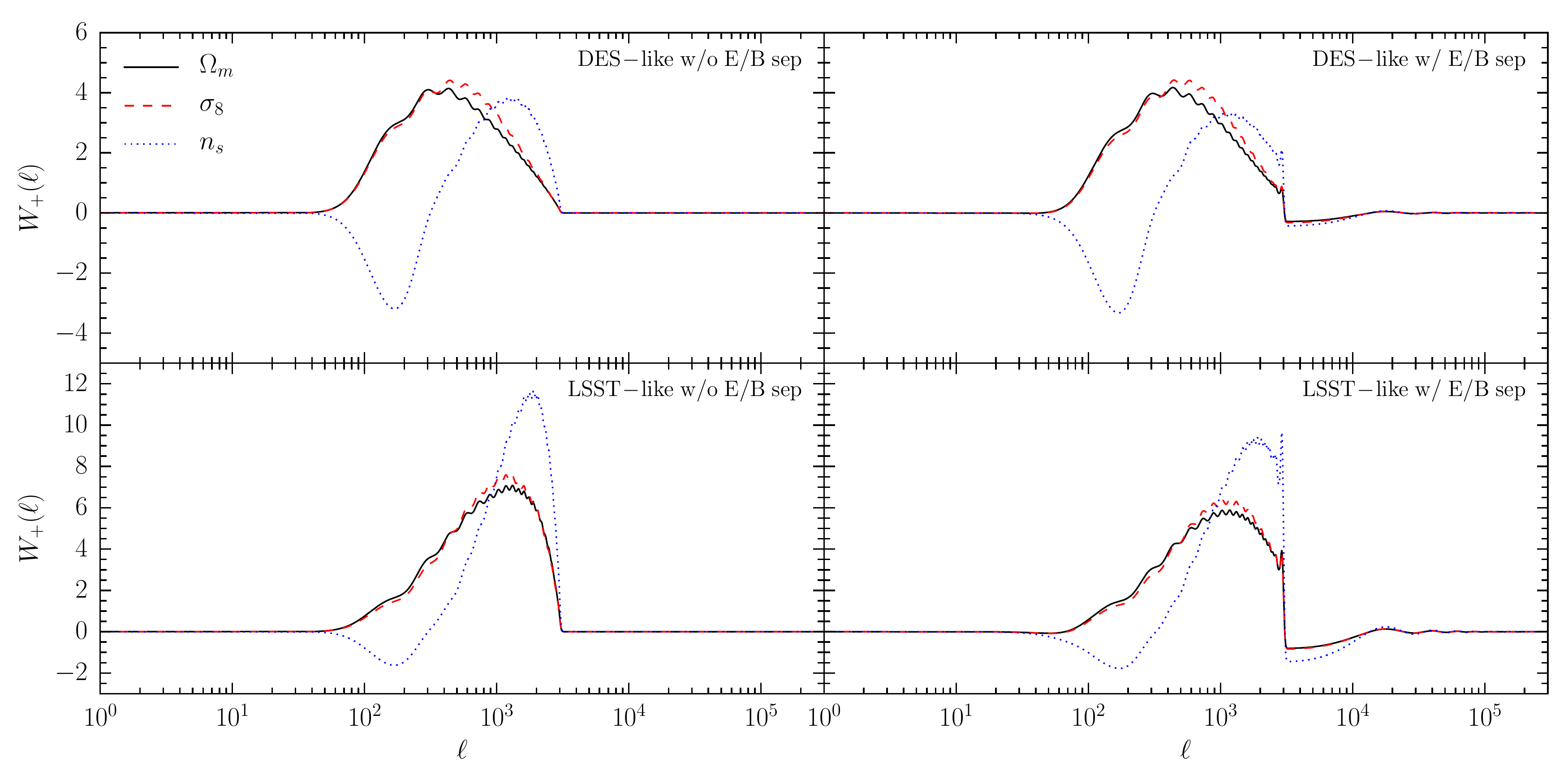}
\end{center}
\caption[]{Band-power window functions $W_{+}(\ell)$ for the optimally compressed statistics for $\Omega_{m}$ (black solid), $\sigma_{8}$ (red dashed) 
and $n_{s}$ (blue dotted). We show results for a DES-like survey in the top panels and an LSST-like survey in the bottom panels. The case of no 
E/B-mode separation is shown in the left column and the case with E/B-mode separation is show in the right column.\label{fig:compwin}}
\end{figure*}

\subsection{Fisher Information Content and Data Compression}\label{sec:dc}
In this section we compute the Fisher information content in the band-power estimators and explore how efficiently these estimators 
can compress the shear correlation function data. We use a $\Lambda$CDM cosmological model with $\Omega_{m}=0.25$, 
$H_{0}=h^{-1}100$ \kms\ Mpc$^{-1}$, $h=0.7$, $\sigma_{8}=0.8$, $n_{s}=1.0$ and $\Omega_{b}=0.044$. We consider only the Gaussian 
and shape noise contributions to the covariance matrix of the shear correlation functions, use the \citet{smith2003} fitting function for the 
non-linear power spectrum, and the \citet{eisenstein1998} fitting formula for the transfer function. We additionally consider both DES- and LSST-like 
surveys with source densities $\bar{n}$ and areas $\Omega_{s}$ of $(\bar{n},\Omega_{s})=(10$ galaxies/arcmin$^{2}$, 5,000 deg$^{2})$ and 
$(\bar{n},\Omega_{s})=(40$ galaxies/arcmin$^{2}$, 20,000 deg$^{2})$ respectively. Finally, we fix all lensing sources to redshift 1. 

We use as a metric for the information content of the survey, $f=\sqrt{|\mathbf{F}|}$, the square root of the determinant of the 
Fisher information matrix. As we are using a single source plane, we consider the information content only in $\Omega_{m}$, $\sigma_{8}$ and 
the spectral index $n_{s}$. See B13 for the details of the Fisher matrix computation. We leave the exploration of data compression methods 
for tomographic lensing analyses for future work. However, note that to the extent that the majority of the extra information in a tomographic 
as compared to a non-tomographic analysis comes from the relative amplitudes of the shear correlation functions or power spectra as a 
function of redshift, we expect results for a single source plane to still be useful since they represent the amount of data compression 
possible in any given tomographic bin. The data compression possible for the entire tomographic analysis can then be estimated by straight 
forward combinatorics, though this estimate should be too pessimistic. These issues have been explored in \citet{asgari2014} where further 
data compression in the case of tomography was possible.

We show the results of our Fisher matrix analysis in Figure~\ref{fig:finfo}. Here we show the amount of Fisher information, characterized by $f$, 
as a function of the number of band-powers used (where the band-powers are ordered by lowest Fourier mode to highest) for both a DES-like 
and LSST-like survey. For comparison, we show the amount of information directly in the shear correlation functions as a function of the smallest 
angular scale considered, from 20 to 5 arcminutes, with the dotted lines. In general, the amount of information increases greatly as smaller scales 
are considered, as is well known in weak lensing \citep[e.g.,][]{takada2009}. The band-power estimators generally have very little weight below $\approx$5-10 
arcminutes. We find that with as few as $\sim10-20$ band-powers, we can recover the majority of the information in the shear correlation functions above 
10 arcminutes. Note that the estimators with E/B-mode separation are less well localized in Fourier space. Their high-$\ell$ tails, even though they exhibit 
oscillations, contribute to the total Fisher information content so that they can give a more efficient data compression at the cost of having less control of 
which scales are used. 

Finally, we consider a nearly optimal data compression scheme, following \citet{tegmark1997}. First, note that the Fisher information matrix can be written as \citep[e.g.,][]{tegmark1997}
\begin{equation}
F_{ij} = \frac{1}{2}\mathrm{Tr}\left[\mathrm{C}^{-1}\left(\vec{\mu}_{,i}\,\vec{\mu}^{T}_{,j} + \vec{\mu}_{,j}\,\vec{\mu}^{T}_{,i}\right)\right]\ .
\end{equation}
where $\mathrm{C}$ is the covariance matrix and $\vec{\mu}_{,i}$ is the derivative of the mean data values with respect to parameter $i$.
For a single parameter, it is clear that the linear combination of data points which contributes most to the trace of the quantity in brackets 
will contain the most information about the parameter under consideration. Thus if we decompose the matrix in the brackets into its eigenvalues and 
eigenvectors, we can perform nearly optimal data compression by retaining only the linear combination of data points given by the eigenvalue and 
eigenvector pair with the largest eigenvalue. For multiple parameters, we choose to simply retain the most efficient linear combination for each parameter 
considered separately. Note that the exact set of optimal linear combinations depends both on the analysis under consideration and the properties of the 
survey, which are all encoded in the data vectors and their covariance matrix. Also, one should be able to compute the optimal set of statistics from approximate 
covariance matrices, like those from the halo model applied to cosmic shear \citep[e.g.,][]{takada2009}, as opposed to covariance matrices computed directly 
from simulations (see also \citeauthor{asgari2014}~\citeyear{asgari2014}). We leave the exploration of these specific issues to future work and instead give below 
an example of potential amount of data compression possible.

We use the process outlined previously to find the three linear combinations of the band-power estimators which retain the most information on 
$\Omega_{m}$, $\sigma_{8}$ and $n_{s}$. The solid point line in Figure~\ref{fig:finfo} shows the amount of information in just these three linear 
combinations. We find that with just these linear combinations of the band-powers, we can retain essentially all of the information in all twenty of the 
band-powers used in this work. In Figure~\ref{fig:compwin}, we show the effective band-power Fourier space windows for the three compressed 
statistics. These are quite similar to those found by \citet{asgari2014}. Note that the window function for $n_{s}$ is both positive and negative as expected. 
We have repeated this analysis with just $\sigma_{8}$ and $\Omega_{m}$ and find that all of the information in the band-powers on these parameters 
can be compressed into just two statistics. Finally, in agreement with \citet{asgari2014}, we find that the amount of data compression is only changed by 
at most a few percent in $f$ when using a different cosmology ($\Omega_{m}=0.28$, $\sigma_{8}=0.84$ and $n_{s}=0.96$, all other parameters are the same) 
to compute the compressed basis linear combinations. 

\section{Conclusions}\label{sec:conc}
In this work, we have derived two new sets of band-power estimators for cosmic shear power spectra. There are several key 
features of these estimators which we expect will make them quite useful for cosmic shear data analysis.
\begin{enumerate}
\item The band-powers can be written and computed as linear combinations of the \textit{binned} real-space 
shear correlation functions and thus take the binning of the correlation functions into account explicitly.
\item The covariance matrix of the band-powers can be derived directly from a linear transformation of the shear correlation 
function covariance matrix. 
\item The estimators account for the survey window function directly in real-space.
\item The estimators do not use shear correlation function data at zero separation and are thus have no biases 
due to shape noise.
\item The window functions of the band-powers in $\ell$-space are compact and largely non-oscillatory, making them easy to 
interpret. This feature also allows one to exclude regions of $\ell$-space where models for the lensing power spectra may have 
significant uncertainties, due say to galaxy formation (see references above).
\item These estimators can be used to construct very efficient data compression schemes with only three band-powers required to capture 
all of the information on the parameters $\Omega_{m}$, $\sigma_{8}$ and $n_{s}$ in the shear correlation functions in the range of 
$\sim10-400$ arcminutes for single tomographic bin.
\end{enumerate}
Importantly, the bin window functions $W_{i}(\theta)$ can be measured from the cosmic shear source galaxy positions 
themselves.\footnote{Note that it is the positions of these galaxies that define the cosmic shear window function. In other words, 
the cosmic shear window function answers the question where did we measure the shear field, not where we could have 
measured the shear field (which is the information carried in the tradition survey mask, but modified for the selection of cosmic 
shear sources).} This measurement is possible because the typical signal-to-noise of cosmic shear for a single galaxy pair is 
$\approx0.01/(0.3\sqrt{2})\approx1/40$ whereas the signal-to-noise of the number of cosmic shear pairs 
as a function of angle (i.e., the unnormalized bin window function) is $1/\sqrt{1}=1$. Thus the signal-to-noise of the bin window 
function measurement for any angular bin will always be $\approx\!40\times$ higher than that of the cosmic shear 
correlation functions. Therefore incredibly low-noise measurements of the bin window functions can be made from the cosmic 
shear catalogs themselves. Given these practical considerations of their implementation and the features listed above, we expect 
the estimators derived in this work to be quite useful for cosmic shear data analysis.

\section*{Acknowledgments}

MRB thanks Bhuvnesh Jain, Tim Eifler, Elisabeth Krause and Sarah Bridle for 
enlightening discussions of cosmic shear window functions. MRB and ER thank Scott Dodelson 
for helping to inspire this work. This work made extensive 
use of the NASA Astrophysics Data System and {\tt arXiv.org} preprint server.

\bibliographystyle{mn2e_good}
\bibliography{refs}

\bsp
\label{lastpage}

\end{document}